\documentclass[journal=jacsat,manuscript=article]{achemso}

\usepackage[version=3]{mhchem} % Formula subscripts using \ce{}
\usepackage{chemformula} % Formula subscripts using \ch{}
\usepackage[T1]{fontenc} % Use modern font encodings
\usepackage{caption}
\usepackage{subcaption}
\usepackage{mathtools}

\usepackage{booktabs}
\author{Zhenghao Wu}
\affiliation{Department of Civil and Environmental Engineering, Northwestern University, 2145 Sheridan Road,
Evanston, Illinois 60208-3109, United States}
\author{Subhadeep Pal}
\affiliation{Department of Civil and Environmental Engineering, Northwestern University, 2145 Sheridan Road,
Evanston, Illinois 60208-3109, United States}
\author{Sinan Keten}
\email{s-keten@northwestern.edu}
\affiliation{Department of Civil and Environmental Engineering, Northwestern University, 2145 Sheridan Road,
Evanston, Illinois 60208-3109, United States}
\alsoaffiliation{Department of Mechanical Engineering, Northwestern University, 2145 Sheridan Road, Evanston,
Illinois 60208-3109, United States}
\title[An \textsf{achemso} demo]
{Implicit Chain Particle Model for Polymer Grafted Nanoparticles}

\abbreviations{IR,NMR,UV}
\keywords{American Chemical Society, \LaTeX}

\begin{document}

\begin{tocentry}

\includegraphics[scale=0.1]{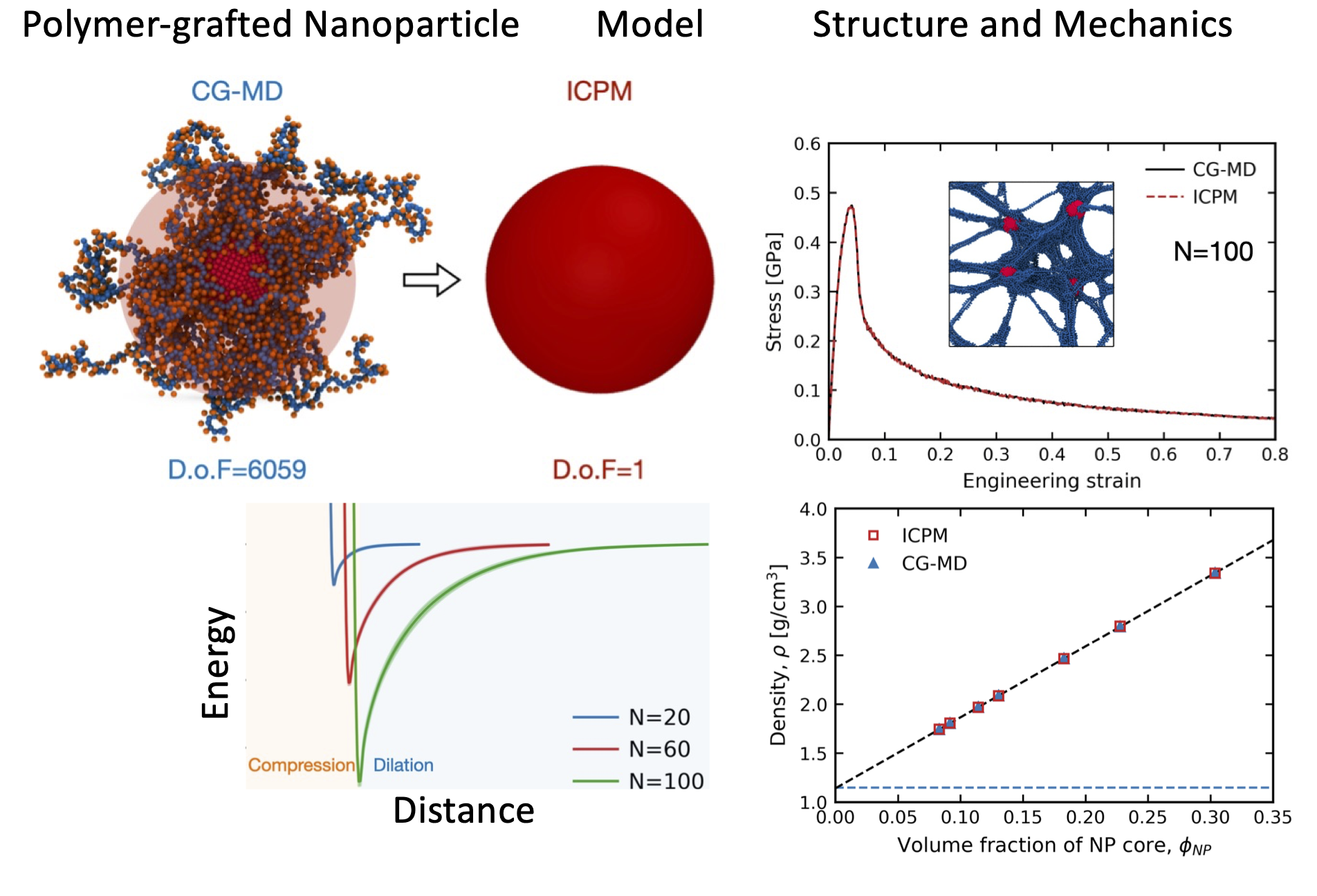}
% shown on the last page

\end{tocentry}

\begin{abstract}

Matrix-free nanocomposites made from polymer grafted nanoparticles (PGN) represent a paradigm shift in materials science because they greatly improve nanoparticle dispersion and offer greater tunability over rheological and mechanical properties in comparison to neat polymers. Utilizing the full potential of PGNs requires a deeper understanding of how polymer graft length, density, and chemistry influence interfacial interactions between particles. There has been great progress in describing these effects with molecular dynamics (MD). However, the limitations of the length and time scales of MD make it prohibitively costly to study systems involving more than a few PGNs, even with bead-spring coarse-grained models. Moreover, it remains unclear how to properly address shortcomings of these models in describing the rate-dependent constitutive response of polymers in a chemistry-specific fashion. Here, we address some of these challenges by proposing a new modeling paradigm for PGNs using a strain-energy mapping framework involving potential of mean force (PMF) calculations. In this approach, each nanoparticle is coarse-grained into a representative particle with chains treated implicitly, namely, the implicit chain particle model (ICPM). Using a chemistry-specific coarse-grained molecular dynamics (CG-MD) model of poly(methyl-methacrylate) as a testbed, we derive the effective interaction between particles arranged in a closed-packed lattice configuration by matching bulk dilation and compression strain energy densities up to failure. We establish an iterative optimization scheme to fine-tune PMFs in the ICPM model to accurately match the stress-strain behaviors during dilation and compression tests. The strain-rate dependence of the mechanical work done is quantified to reveal that the interparticle potential can be expressed with a strain rate dependent energy well depth that culminates in a simple power law Cowper-Symonds strain hardening model. Given the aggressive degree of coarse-graining ($\sim 1:10000$) involved, the scope and limitations of the ICPM model are cautiously discussed. Overall, the ICPM model increases the computational speed by approximately $5-6$ orders of magnitude compared to the CG-MD models. This novel framework is foundational for particle-based simulations of PGNs and their blends and accelerates the understanding and predictions of emergent properties of PGN materials.

\end{abstract}

\section{Introduction}
 
Incorporation of inorganic nanoparticles into polymers has been extensively utilized to produce nanocomposites that possess enhanced mechanical, thermal, or electrical properties relative to neat polymers\cite{balazs_nanoparticle_2006,kumar_nanocomposites_2010,kumar_nanocomposites_2013,kumar_50th_2017}. These improvements usually require a homogeneous dispersion of nanoparticles in the polymer matrix. However, conventional inorganic nanoparticles are commonly incompatible with organic polymers, which induces aggregation of nanoparticles or even total phase separation. Such inhomogeneities are usually considered to lead to degradation of material performance because the advantages of a high surface area to volume ratio of nanoparticles (hence formation of interphases) cannot be realized. To overcome these issues, grafting polymer chains on the surface of a nanoparticle is emerging as a useful technique to increase the compatibility between inorganic nanoparticles and organic polymers\cite{kumar_nanocomposites_2013}. A particularly promising case is matrix-free polymer grafted nanoparticles (PGNs) \cite{fernandes_hairy_2013}, where PGNs self-assemble into films or composites without the use of ungrafted matrix polymers. Recent experimental studies have identified several unique features of matrix-free PGN assemblies compared with the traditional polymer nanocomposites. These include the ability to form very stiff films with exceptional ordering of nanoparticles, the formation of amorphous films with extraordinary slow relaxations governed by colloidal jamming, superior impact resistance at high-strain rates due to ductile unravelling of grafted chains, and unusual transport properties due to interfacial relaxation mechanisms near nanoparticle surfaces and graft interfaces.\cite{akcora_anisotropic_2009,bilchak_tuning_2020,parisi_universal_2021,jhalaria_unusual_2022}. Recent advances in experiments have extensively examined various material properties of matrix-free PGN composites such as chain conformations\cite{chevigny_polymer-grafted-nanoparticles_2011}, dynamics\cite{miller_simulation_2022}, glass transition\cite{askar_polystyrene-grafted_2017}, and mechanical properties\cite{jhalaria_unusual_2022}. However, understanding the molecular mechanisms underpinning these emergent behaviors remains extremely challenging with experimental techniques. Particularly for mechanical response, the predominant molecular mechanisms at the root of rheological observations, the extracted relaxation times, and failure progression especially in the high-frequency or high-strain rate regime at glassy state are difficult to ascertain. Molecular dynamics (MD) simulations have emerged as a powerful complementary tool for this purpose because they reveal local failure mechanisms with a monomer-level description of conformations over dynamic trajectories with femtosecond resolution. This has facilitated a deeper understanding of the relationship between molecular design parameters, e.g., polymer graft length, density and chemistry, and macroscopic material properties for optimal design of matrix-free PGN materials\cite{hansoge_materials_2018,hansoge_universal_2021,midya_structure_2020}.

All-atom molecular dynamics (AA-MD) simulations are able to address some of the aforementioned tasks, but their practical application in modeling polymer nanocomposites is still limited because of the prohibitive computational cost of this method. For example, AA-MD simulations fall short of predicting self-assembly processes, and even the relaxed configurations of these complex systems are very difficult to obtain in the time frame of AA-MD simulations because of the slow nature of the collective motions pertaining to polymer segments, grafted chains, and nanoparticles. This motivates the development of systematic coarse-grained molecular dynamics  (CG-MD)\cite{mller-plathe_coarsegraining_2002} formulations, which extend the length and time scales accessible to MD simulations by lumping groups of atoms into coarse beads that have effective interactions derived from AA-MD.  Although CG-MD has been extensively used to study the structures, chain conformations, and mechanical properties of matrix-free PGNs\cite{hansoge_materials_2018,ethier_structure_2018,hansoge_effect_2019,ethier_uniaxial_2019}, the spatiotemporal scales accessible by CG-MD still pose a number of limitations to current understanding. First, predicting the large-scale self-assembly or structure evolution of matrix-free PGN materials is difficult due to kinetic traps and sampling challenges. Second, slow colloidal relaxations are very difficult to quantify and access. Third, predicting fracture processes, for instance, for thin films under impact, involves modeling of the process zone reaching micrometer dimensions. \cite{yun_symmetry_2019,parisi_universal_2021,chen_controlling_2022,alkhodairi_fracture_2022}. This calls for the development of accurate and efficient upscaling approaches that enable computational modeling of matrix-free PGNs at even larger length and time scales.
%However, the length and time scales accessible by CG-MD simulations are still limited to tens of nanometers and hundreds of nanoseconds, which does not allow to simulate behaviors associated with high-strain rate fractures of polymer nanocomposites. This includes the following issues: (1) one cannot get well-relaxed structures, or do structure prediction (2) one cannot access slow colloidal relaxations (3) one cannot predict fracture of  thin films or model a process zone of micron dimensions.

There are indeed computational methods developed for polymeric systems at mesoscopic scales. Recent studies have used polymer density functional theory\cite{koski_fluctuation_2018}, self-consistent field theory\cite{sides_hybrid_2006} and polymer reference interaction site model\cite{jayaraman_effective_2009,martin_using_2016} to study polymer (grafted) nanoparticles with large system sizes. Although these methods can study properties such as polymer structure and nanoparticle self-assembly, the dynamics and mechanical properties of these materials are often intractable. Moreover, these methods are mostly field-based and rely on the mean-field approximation, whereas having particle-based resolution allows for quantifying particle correlations and thermal fluctuations, which are prominent for studying local dynamics or phase transitions. Additionally, models that can capture large-deformation and fracture behavior of PGNs and their transition from brittle to ductile failure as chain length increases \cite{schmitt2016processing,chen_controlling_2022} are critically needed. Therefore, an efficient CG model that can predict the thermomechanical properties of PGNs while reaching length scales above $\sim 100$ nm remains to be established. One way to achieve this is to develop a particle-based model \cite{yin2022simplified,li2002meshfree} with even more aggressive coarse-graining using effective potentials between NPs to implicitly model the polymer grafts in PGNs.

As a step toward addressing this problem, the effective interaction between two planar surfaces with grafted polymer brushes was shown to be well captured by an effective potential of mean force (PMF) in recent work by our group \cite{hansoge_universal_2021}. In that study, the effective repulsive and attractive potentials were derived from compression and tensile simulations of chemistry-specific CG-MD polymer models using a strain-energy conservation approach. The advantage of this method is that the work associated with the unraveling of interacting chains is well captured, which lands itself well to mechanical property predictions, especially in the large-strain, large strain-rate regime. The effective interactions were found to follow a universal relationship with respect to molecular design parameters such as the molecular weight of polymer chains, the graft density, and the monomer chemistry. Moreover, this model captured the strain-rate dependence of effective interactions with a Cowper-Symonds type power law for the total interaction energy. These features set our potential-of-mean-force approach apart from most existing methods, e.g., where the pairwise PMF is estimated from the interaction of two or three graft nanoparticles by manually varying their spacing, and the computed PMFs are usually not able to capture the cohesive interactions well, showing purely repulsive potentials\cite{estridge_effect_2015,munao_molecular_2018}. Here, we extend the development of effective interactions between two polymer grafted slabs to pairwise interactions between NPs arranged in a closed-pack unit cell to simulate a complete matrix-free PGN system at the micrometer scale. We demonstrate that the effective pairwise interactions between two spherical PGNs can be derived from PMFs evaluated by varying their radial distance under dilation/compression deformations of a unit cell in our chemistry-specific CG-MD simulations. The derived effective interparticle potentials are then utilized to perform particle-based MD simulations of matrix-free PGNs with implicit polymer chains, which is called the ICPM model. The stress-strain curves from the dilation and compression in the ICPM models agree well with those from the corresponding fine-grained model. Additionally, an iterative optimization scheme is proposed to fine-tune the PMF in the ICPM model to accurately match the stress-strain behavior during dilation and compression loadings. Finally, we show that the ICPM model significantly accelerates large-scale simulations of PGNs by extending the spatiotemporal scale by 5-6 orders of magnitude compared to classic CG-MD models. The novel framework developed here is expected to accelerate the characterization, understanding, and predictions of the structures, phase behavior, and thermomechanical properties of matrix-free PGNs and their blends.

\section{Methodology}

\begin{figure}
    \centering
    \includegraphics[width=0.7\textwidth]{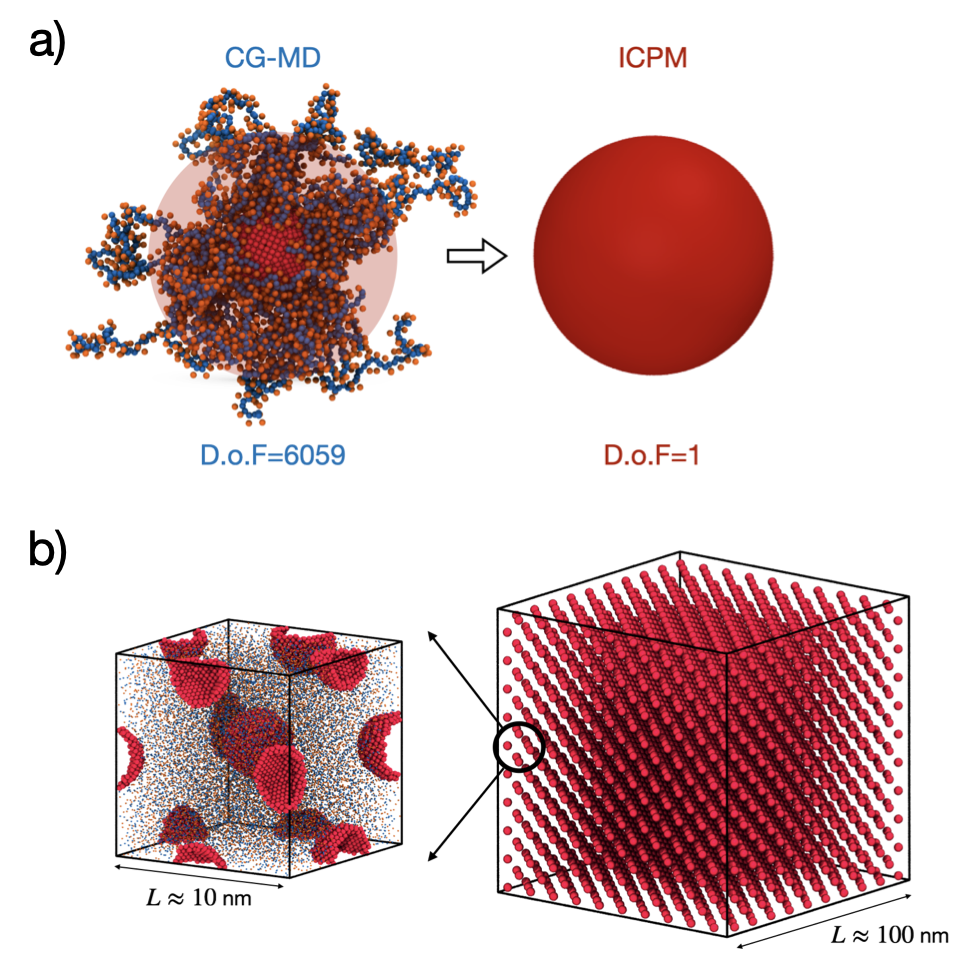}
    \caption{(a) Coarse-grained mapping scheme from a CG-MD model to a ICPM of a single PMMA-PGN with graft length N=100 monomers; Blue, orange, and red beads represent the polymer backbone, side groups and nanoparticle surface beads, respectively. Accordingly, the degree of freedom decreases from 6059 (beads) in CG-MD to 1 (bead) in ICPM. (b) a unit cell of matrix-free PGNs with graft length N=100 monomers in the CG-MD representation and a simulation box comprises multiple unit cells in the ICPM representation.}
    \label{fig:CG_mapping}
\end{figure}

A hierarchical coarse-graining framework is developed to derive the effective interparticle potential, which is a key to our approach. The construction of the hierarchical simulation methodology consists of three components: all-atom (AA-MD), coarse-grained (CG-MD), and implicit chain particle model (ICPM) which is the highest level of coarse-graining in this study. A mapping scheme of an ICPM to a CG-MD representation of PGNs is shown in Figure \ref{fig:CG_mapping}. In this contribution, the major goal is to demonstrate the protocol for deriving PMF between PGNs based on the CG-MD model and using these PMFs as effective interparticle interactions to replace explicit polymer grafts in the ICPM model. The ICPM model is developed to predict thermomechanical properties consistent with the underlying fine-grained model. In the following subsections, we will first introduce the reference fine-grained model. Subsequently, the theoretical foundation and derivation procedure of the effective interactions, i.e., the PMFs, between PGNs will be presented. Finally, we will demonstrate an iterative optimization algorithm to fine-tune the interparticle PMF to accurately reproduce the mechanical properties of the reference CG-MD model.

\subsection{Coarse-Grained (CG) Model}

In this work, a temperature-transferable systematic CG-MD model of poly(methyl methacrylate) (PMMA)\cite{hsu_systematic_2014} serves as a fine-grained reference model for the derivation of chemistry-specific PMFs. In this model, a repeating unit (monomer) is mapped onto two CG beads that are placed in the center of mass of the methacrylate backbone group and a side-chain group. This CG model of polymers retains the thermodynamic, dynamics, and mechanical properties of the underlying AA-MD model, thus providing a firm basis for the development of a thermomechanically consistent highly CG model. Here, we refer the reader to previous papers for details of this coarse-graining technique\cite{xia_energy-renormalization_2017,xia_energy_2019}. The nanoparticle (NP) is also represented by a CG model. In the NP model, a shell consisting of multiple beads is arranged in an FCC lattice in which the nearest neighbors and next-nearest neighbors interact through harmonic bonds. The mass of the bead can be determined based on the density of the nanoparticle to be modeled; in this case, we choose the mass of the bead $162$ g / mol, similar to the value used in our previous work\cite{hansoge_universal_2021}. Indeed, such a CG model for NPs has several advantages. First, the network structure makes the CG model of NPs quite stiff, similar to that of inorganic particles, and they do not exhibit any appreciable deformation during mechanical tests. Next, explicit beads on the surface of NP introduce surface roughness and friction, making our NP model more realistic for describing interfacial interactions between nanoparticles and polymers\cite{hanakata_unifying_2015}. In addition, the shell configuration (without modeling the inner beads of nanoparticles) further reduces the computational cost. To model PGNs, we use covalent bonds to graft polymer chains onto the surface beads of NP, with the grafting sites evenly distributed on the NP surface. To demonstrate the development of the ICPM method, we use PMMA-PGNs with various graft chain lengths ranging from 20 to 100 monomers at a fixed graft density $0.5$ $\mathrm{chains/nm^2}$ as reference fine-grained systems. Details of the force fields used for polymers and nanoparticles and the equilibrium graft-chain conformations are given in \textbf{Supporting Information}. It is noted that the relatively generic nature of the NP model gives an opportunity to systematically study how the strength of the interfacial interaction and the stiffness of NPs\cite{zhu2022effect} affect the mechanical properties of matrix-free PGNs,  which is left to be pursued in future work.   

\subsection{Theory for Derivation of Effective Interparticle Potentials}

\begin{figure}
    \centering
    \includegraphics[width=1.0\textwidth]{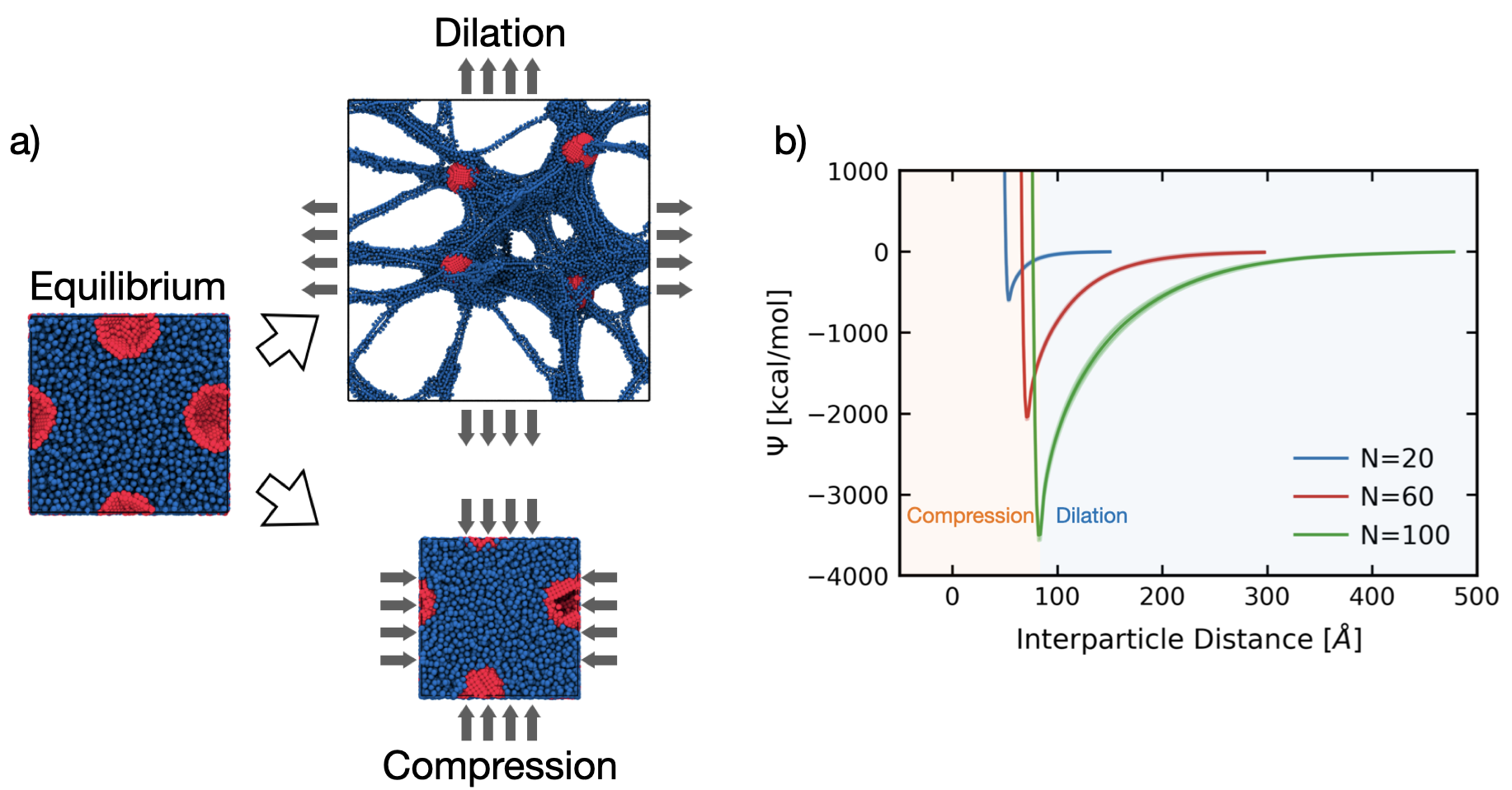}
    \caption{(a) schematic of dilation/compression deformations from PMMA PGN melts at equilibrium in the CG-MD representation; (b) Examples of potentials of mean force (PMF) as a function of interparticle distances derived from CG-MD models of PGN melts with graft chain length $N=20$ monomers (blue), $N=60$ monomers, and $N=100$ monomers; The repulsive and attractive potential from the compression and dilations deformations are separated by light orange and blue backgrounds, respectively, for PGN melts with $N=100$ monomer.}
    \label{fig:PMF_derive}
\end{figure}

We derive the effective interaction between PGNs using the so-called strain-energy mapping approach\cite{ruiz_coarse-grained_2015,hansoge_universal_2021}. Here, the effective interaction between particles is termed the PMF, which is basically the free energy along a certain reaction coordinate. The PMF essentially removes all degrees of freedom except for the reaction coordinate to describe a diffusive motion of the system within the effective potential. In the strain-energy mapping, non-equilibrium molecular dynamics (NEMD) simulation is performed to induce the free-energy perturbation of the system. In principle, the free-energy deviation $\Delta F$ can be determined from the statistical distribution function in equilibrium simulations. However, it is usually difficult to estimate using this forward method due to the sampling efficiency. In contrast to attempting to obtain well-sampled statistical distributions, we connect the non-equilibrium process to the equilibrium property of PMF considering Jarzynski's non-equilibrium work relation\cite{jarzynski_nonequilibrium_1997}, in which the deviation of free energy $\Delta F$ is related to the amount of external work $W$ performed on a system. The amount of external work $W$ during NEMD can be estimated using the strain energy calculation formula in continuum mechanics:
\begin{equation}
    \Delta F \approx W =V\times \mathcal{W}(\epsilon)=V\times \sum\limits_{x,y,z}\int\limits_0^{\epsilon}\sigma d \epsilon
    \label{equ:work_free_energy}
\end{equation}
where $V$ is the volume of the simulation box; $\mathcal{W}(\epsilon)$ denotes the strain energy density; $\sigma$ and $\epsilon$ represent the stress and the engineering stress, respectively. In practice, running more NEMD trials will reduce the error of $W(\epsilon)$ and then in $\Delta F$ but increase the computational cost. As mentioned at the beginning of the section, the PMF $\Psi(\xi)$ can then be determined along a selected collective coordinate $\xi$ based on $\Delta F$. 

In theory, PMF $\Psi(\xi)$ is a many-body interaction that includes complicated correlations of the local environment due to coarse-graining. Here, as a first attempt to achieve this aggressive coarse-graining in ICPM, we use additive pairwise interactions to represent the potential energy surface for simplicity. Accordingly, we choose the distance between particles $r$ as our collective coordinate $\xi=r$, giving the following:
\begin{equation}
    \Delta F(\mathbf{R})=\sum\limits_{i,j} \Psi(r_{ij})
    %V_{\mathrm{PMF}}(r)=\frac{U(\epsilon(r))}{\Omega}
    \label{equ:U_pair}
\end{equation}
where $\mathbf{R}$ represents the coordinates of the particles in the system and $\sum$ denotes the sum of interactions with neighbors. In our approach, we perform dilation/compression as shown in Figure \ref{fig:PMF_derive} (a) to generate conformational changes and therefore the free energy deviation $\Delta F(\epsilon)$ as a function of the engineering strain $\epsilon$ of Equation \ref{equ:work_free_energy}. Thus, a transformation function is required to convert the engineering strain $\epsilon$ to the distance between particles $r$ to build the pairwise PMF $U(r)$. We employ a simulation setup of particles arranged in a face-centered cubic (FCC) unit cell that consists of four PMMA-grafted nanoparticles in the CG-MD representation, as illustrated in Figure \ref{fig:CG_mapping}. Under the condition $T<T_g$ where the mobility of PGNs is limited, the FCC configuration of PGNs is well preserved during the simulation. This closed-packed lattice configuration provides a means of projecting the volumetric change during deformation at a certain engineering strain $\epsilon$ to the interparticle distance. This results in a rough estimate of the distance between particles based on engineering strain: $r=L_0\times(1+\epsilon)/\sqrt{2}$, where $L_0$ is the edge size of a cubic simulation box in equilibrium. It is noted that various types of superlattice structures are experimentally observed in matrix-free polymer (ligand) grafted nanocrystals, depending on the polymer graft length, density, chemistry, and core size\cite{ye_structural_2015}.

\begin{figure}
    \centering
    \includegraphics[width=0.6\textwidth]{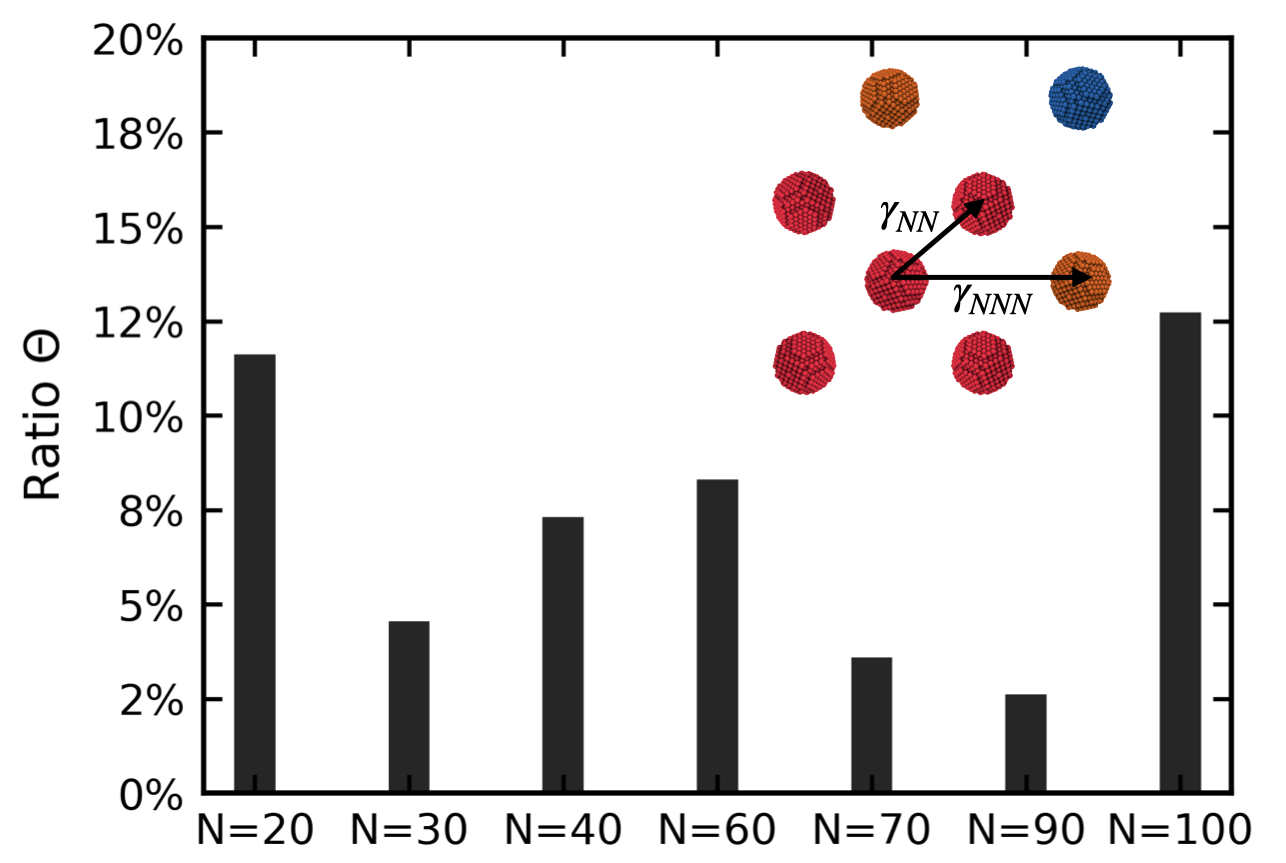}
    \caption{Ratios $\Theta$ of cohesive interaction energy from next-nearest neighbors $\gamma_{pp}^{NNN}$ and nearest neighbors $\gamma_{pp}^{NN}$ estimated from CG-MD models of matrix-free PGNs as a function of graft chain lengths. Inset: Schematic to explain the nearest neighbors $\gamma_{NN}$ and next nearest neighbors $\gamma_{NNN}$ in a FCC structure. }
    \label{fig:ratio_NN_NNN}
\end{figure}

\subsection{Implicit Chain Particle Model}

The implicit chain particle model (ICPM) is a coarse-grained particle-based model for matrix-free PGNs, in which point particles interact with PMFs that are systematically derived from fine-grained models with explicit grafts. In practice, we consider the nearest-neighbor interactions between PGNs in the current ICPM model. This approximation is indeed supported by the quantitative analysis of the cohesive interactions ($\gamma_{pp}$) contributed by the nearest neighbors (NN) and the next-nearest neighbors (NNN), as shown in the inset of Figure \ref{fig:ratio_NN_NNN}. Figure \ref{fig:ratio_NN_NNN} displays the ratios $\Theta$ of the cohesive interaction energy of the next-nearest neighbors (NNN) and nearest neighbors (NN) estimated from CG-MD models as a function of graft chain length. The cohesive interaction energy is quantified by the summation of pairwise non-bonded interactions between all particles in respective pairs of PGNs: 
\begin{equation}
    \Theta=\frac{\gamma_{pp}^{NNN}}{\gamma_{pp}^{NN}}
    \label{equ:V_pair}
\end{equation}
%=\frac{\sum\limits_{i,j}^{N_{NNN}} 4\epsilon\bigg(\big(\frac{\sigma}{r_{ij}}\big)^{12}-\big(\frac{\sigma}{r_{ij}}\big)^{6}\bigg)}{}
where $\gamma_{pp}=\sum\limits^{i\in N_p, j\in N_q} 4\epsilon\bigg(\big(\frac{\sigma}{r_{ij}}\big)^{12}-\big(\frac{\sigma}{r_{ij}}\big)^{6}\bigg)$ with $\sigma$ and $\epsilon$ being force-field parameters for polymers that can be found in \textbf{Supporting Information}; $N_p$ and $N_q$ are the numbers of polymer beads in the PGN $p$ and $q$, respectively. The ratio for all graft chain lengths is $\Theta<15\%$, which is in agreement with our assumption of short-range NN interactions. Based on this analysis, we derive the PMF from a FCC unit cell of matrix-free PGNs, leading to a simple relation between $\Delta F(\mathbf{R})$ and the pair-wise effective interactions $\Omega_{ij}(r)$:
\begin{equation}
    \Delta F(\mathbf{R})=\frac{\sum\limits_{i,j}^{\Omega}\Psi_{ij}(r)}{\Omega}
    \label{equ:V_pairDF}
\end{equation}
where $\Omega$ is a normalization factor that is essentially the total number of pair-wise interactions. Specifically, in an FCC unit cell with periodic boundary condition applied on the x, y, and z axis, there are 4 particles, each interacting with 12 nearest neighbors and another 6 next-nearest neighbors. It results in a total of $\mathcal{N}_{NN}=4\times12=48$ pairs of interactions for the nearest neighbors and $\mathcal{N}_{NNN}=4\times 6=24$ pairs of interactions for the next-nearest neighbors. Since only the nearest-neighbor interactions are considered, $\Omega=48/2=24$, where the division of 2 is due to the double counting of pair-wise interactions. 

Examples of PMFs derived for PMMA-PGN with graft chain length N = 20, N = 60 and N = 100 monomers are presented in Figure \ref{fig:PMF_derive} (b). The resulting PMFs are obviously chain-length dependent (N-dependence). %In addition, the well depth of the potential of mean force and the equilibrium distance between particles increase with increasing graft chain length.
Moreover, the obtained PMFs have a gradual decay in the attractive region, which involves the intermingling and relaxation process of the chains. This range of this decay extends further in longer grafts. It is critical to note that this long-ranged nature of this potential exceeds the range of NN interactions, which would conflict our initial assumption of NN interactions in ICPM simulations and result in stronger cohesion than desired. For equilibrium states, we can mitigate this issue by shortening the range of PMF to $R_{cut}=1.4\times R_{NN}$ in the ICPM model, where $R_{NN}$ is the equilibrium distance between PGNs, i.e., the distance between the nearest neighbors. This choice ensures that the particles interact with their nearest neighbors only. Under large-strain deformation, particles can rearrange and potentially enter within this cutoff range in ICPM runs. This essentially requires the decoupling between the interaction range and the PMF range. Specifically, a static neighbor list that is initially constructed in equilibrium state with a cut-off distance $R_{cut}=1.4\times R_{NN}$ is employed in the ICPM simulation during deformation tests. Long-range interactions are computed only for particles in the original neighbor list. This is reasonable for the particular scenarios considered here, considering that the structure of polymeric systems far below the glass transition temperature does not vary rapidly and original neighbors are maintained during deformation. %Similar choices are made in simulations of solids and metals\cite{hou_efficient_2013} where particle rearrangements and large-scale flow are not anticipated. 
More technical details of the ICPM simulations are given in \textbf{Supporting Information}.

\subsection{Optimization of Potential of Mean Force}

\begin{figure}
    \centering
    \includegraphics[width=0.5\textwidth]{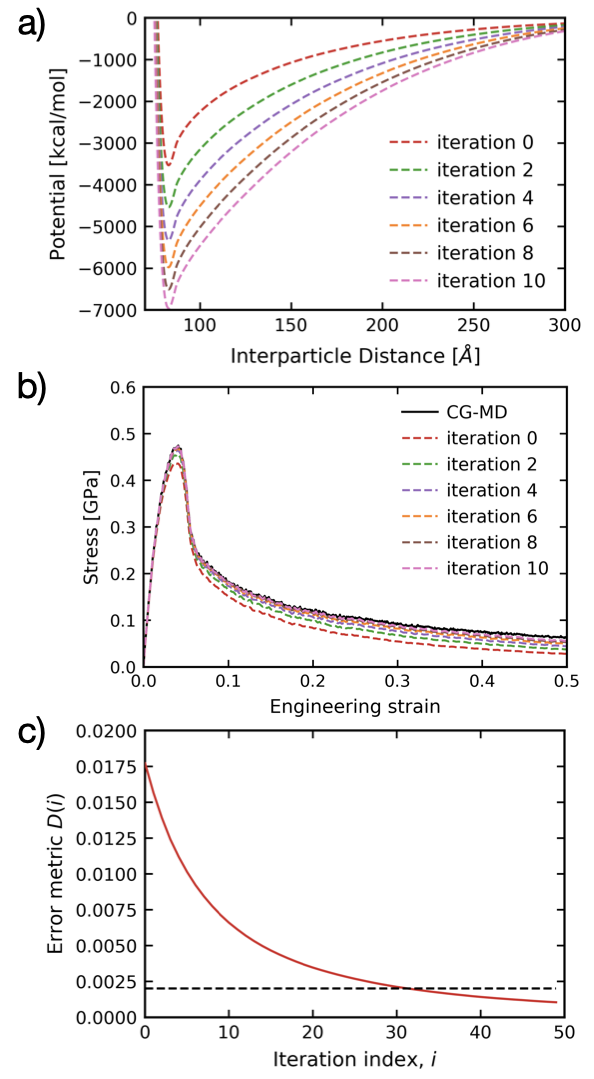}
    \caption{Examples to illustrate the iterative optimization (PMMA grafts with $N=100$ monomers): a) PMF as a function interparticle distance from iterations 0-10; b) stress-strain curves from the ICPM models using the corresponding optimized PMF; the solid line is the reference data from CG-MD simulations; c) Example of PMF optimization for PGNs with PMMA grafts $N=100$ monomers  to illustrate the metric $D$ evolving as increasing iteration index $i$. The dashed black line is the value of the criteria we used to stop the optimization $D(i)=2\times 10^{-3}$ GPa.}
    \label{fig:pot_opt}
\end{figure}

The PMF derived directly from the CG-MD simulation is indeed a free energy instead of a potential energy. Reducing the complex many-body PMF to additive pairwise interactions between nearest neighbors introduces additional errors in the derivation process. Moreover, the point representation of the entire PGN neglects rotational degrees of freedom. Although it has already been able to serve as a good initial estimate of effective pair interactions between PGNs, as demonstrated in Figures \ref{fig:pot_opt} (a) and (b), the PMF can be further refined in an iterative way by adding a correction term:
\begin{equation}
    \Psi(r)^{n+1}=\Psi(r)^n+ \Delta \Psi^n(r)
\end{equation}
and
\begin{equation}
    \Delta \Psi(r)=\frac{\Delta W}{\Omega}=\frac{V\times \Delta{\mathcal{W}(\epsilon)}}{\Omega}
\end{equation}
where $\Psi(r)^{n}$ represents the PMF at iteration $n$; $\Delta{\mathcal{W}(\epsilon)}$ is the deviation of the strain energy density:
\begin{equation}
    \Delta{\mathcal{W}(\epsilon)}=\sum\limits_{x,y,z}\int\limits_0^{\epsilon}(\sigma^{ref}-\sigma) d \epsilon
\end{equation}

with $\sigma^{ref}$ and $\sigma$ being the stress of the underlying fine-grained simulation and the current ICPM simulation at the same strain, respectively. As an example, in Figure \ref{fig:pot_opt} (a), we show the optimization of PMF from iteration $0$ (naive PMF derived directly from CG-MD simulations) to iteration $10$ for PMMA-PGN with graft chain length $N=100$, which is the case that needs the most iterations to reach the stop criterion (details given below). The corresponding stress-strain curves of the reference CG-MD during iterative optimization are also shown in Figure \ref{fig:pot_opt} (b). Clearly, the ICPM model predicts a lower maximum stress, as well as a noticable lower stress after the yield point, compared to the reference CG-MD predictions. 

To quantify the deviation between ICPM and CG-MD predictions of stress-strain curves during iterative optimization, we use the following metric:
\begin{equation} 
   D=\frac{1}{n}\sum\limits^{n}|\sigma^{ref}(\epsilon)-\sigma(\epsilon)|
\end{equation}
where $n$ is the number of data points collected during simulations. This metric is used as the criterion to indicate the convergence of the optimized PMF. In this work, we stop iterative optimization when $D\leq 2\times 10^{-3}$ GPa. For glassy polymers and PGNs studied here, this error in the modulus is less than 1$\%$. Figure \ref{fig:pot_opt} (c) shows the evolution of the metric $D(i)$ as the optimization step $i$ of PGN with PMMA grafts $N=100$ monomers. Generally, approximately 20-40 iterations are required for PMMA-PGN (e.g., $\sim 33$ iterations for $N=100$ monomers) to complete the optimization.

\section{Results}

\subsection{Bulk Dilation and Compression}

\begin{figure}
    \centering
    \includegraphics[width=1.0\textwidth]{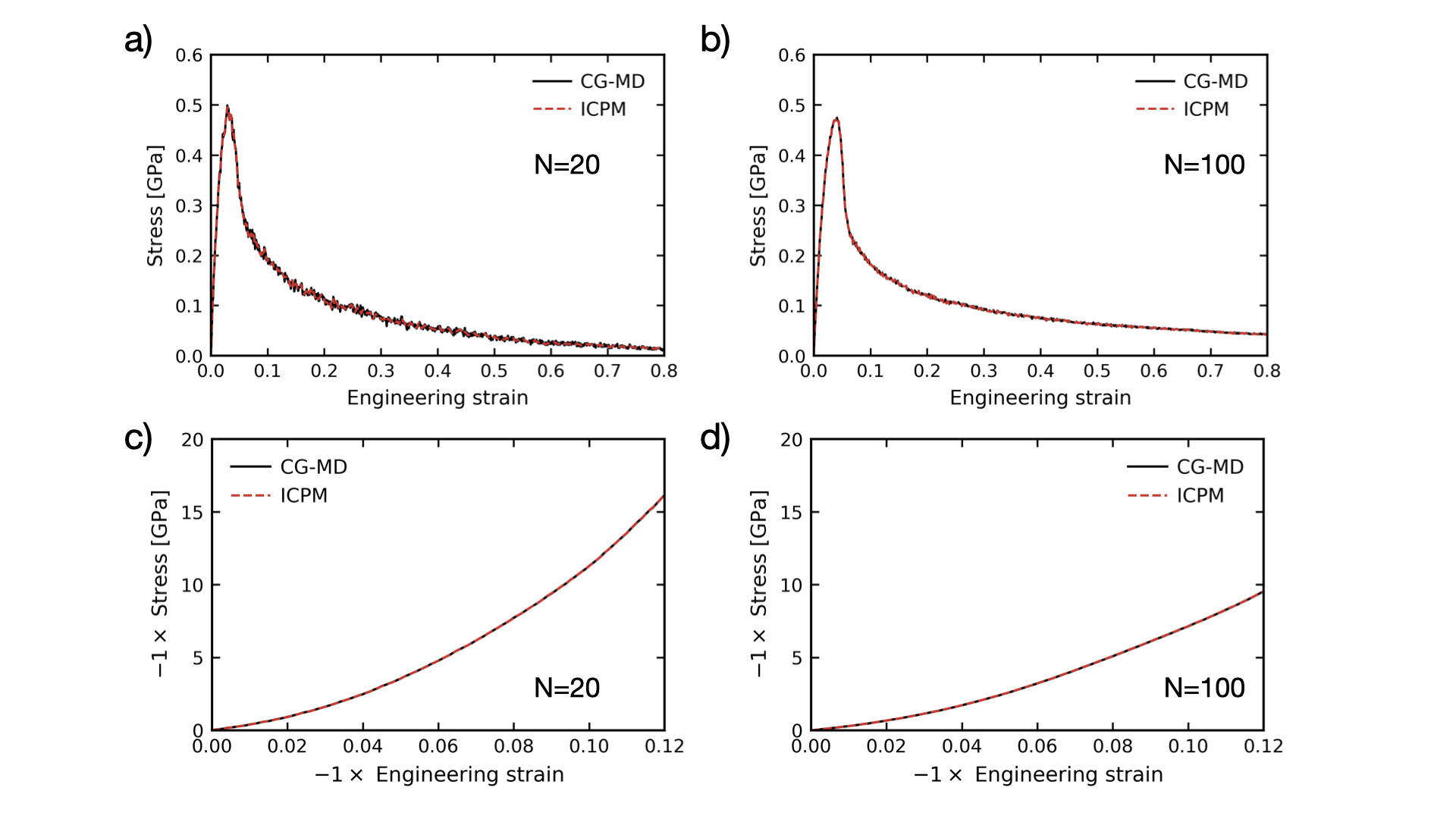}
    \caption{Stress-strain curves from triaxial tensile (a,b) and compression (c,d) deformations of CG-MD (black dashed lines) and ICPM models (red solid lines) of FCC structures of PGNs with PMMA grafts a),c) N=20 monomers and b),d) N=100 monomers.}
    \label{fig:triaxial_deform_stress_modulus_toughness}
\end{figure}

\begin{figure}
    \centering
    \includegraphics[width=1.0\textwidth]{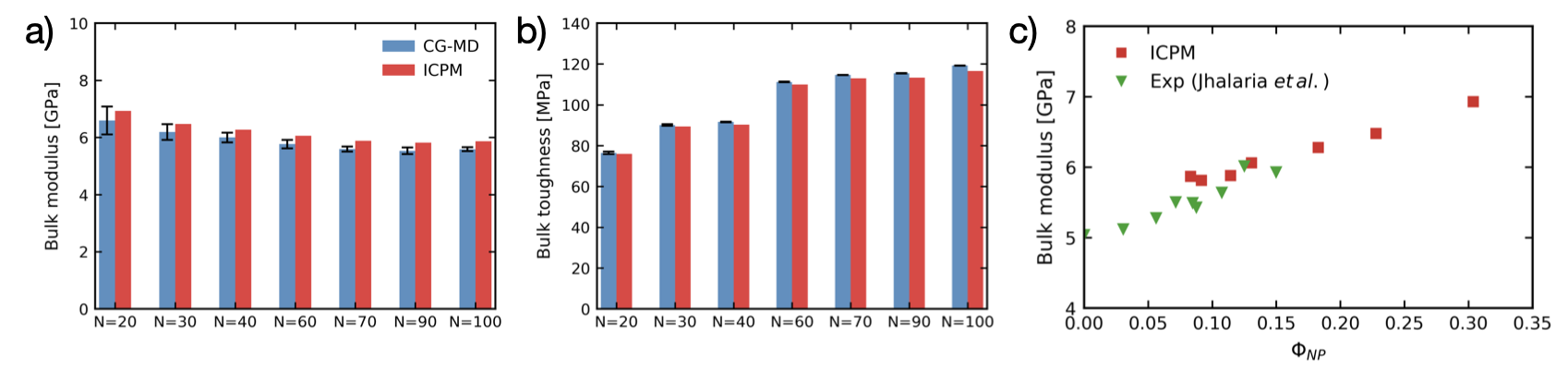}
    \caption{Bulk modulus (a) and toughness (b) estimated from dilation deformation of CG-MD (blue) and ICPM models (red) of FCC structures of PGNs with various PMMA graft lengths; The error bars in CG-MD simulations are standard deviations from 9 trials of CG-MD systems with different initial coordinates and velocities; (c) Comparison of bulk modulus of matrix-free PGNs as a function of volume fractions of nanoparticle cores from ICPM models (red squares) and experimental measurements (green triangles) by Jhalaria \textit{et al.}\cite{jhalaria_unusual_2022}.}
    \label{fig:bulk_modulus_toughness}
\end{figure}

We first validate the ICPM model for predictions of mechanical properties by comparing with results from CG-MD simulations. The stress-strain curves from dilation and compression deformations of the ICPM and CG-MD models of matrix-free PGNs in the FCC structure with short ($N=20$) and long ($N=100$) grafts are shown in Figure \ref{fig:triaxial_deform_stress_modulus_toughness}, respectively. The ICPM results agree well with those of the CG-MD model in a wide range of engineering strains and graft chain lengths. The bulk modulus and toughness as a function of graft chain length $N$ from the corresponding stress-strain behaviors are presented in Figures \ref{fig:bulk_modulus_toughness} a) and b), respectively. Specifically, the bulk modulus is defined as $K=\sigma(\epsilon)/(\delta V/V)$, which is the ratio between stress $\sigma$ and volumetric strain $(\delta V/V)$ up to engineering strain $\epsilon=2\%$ in dilation deformation. The bulk modulus $K$ is found to decrease with increasing graft chain length in our simulations. This is because the volume fraction of the stiff NP core decreases as the graft length increases at a fixed graft density. In addition, the toughness is determined by integrating the entire stress-strain curve. The computed toughness increases with increasing graft chain length, N. This observation is qualitatively aligned with the N-dependence of pure bulk polymer melts. Overall, the ICPM model predicts both modulus and toughness that are in good agreement with the finer-grained CG-MD models at the respective graft lengths. Furthermore, Figure \ref{fig:bulk_modulus_toughness} (c) shows a comparison of the bulk modulus of matrix-free PGNs as a function of the volume fractions of the NP cores from ICPM simulations and a recent experimental measurement\cite{jhalaria_unusual_2022}. The experimental data is based on poly-methacyrlate (PMA) in the glassy high-frequency regime, at comparable volume fractions and employing larger (8 nm, as opposed to 2 nm) silica NPs. As such, one-to-one comparison may be challenging as the degree of confinement of the grafts is influenced by NP size and radius of curvature. Nevertheless, the predictions of ICPM models in the examined region are still in the vicinity of the experimental measurements.

\subsection{Uniaxial Deformation}

\begin{figure}
    \centering
    \includegraphics[width=1.0\textwidth]{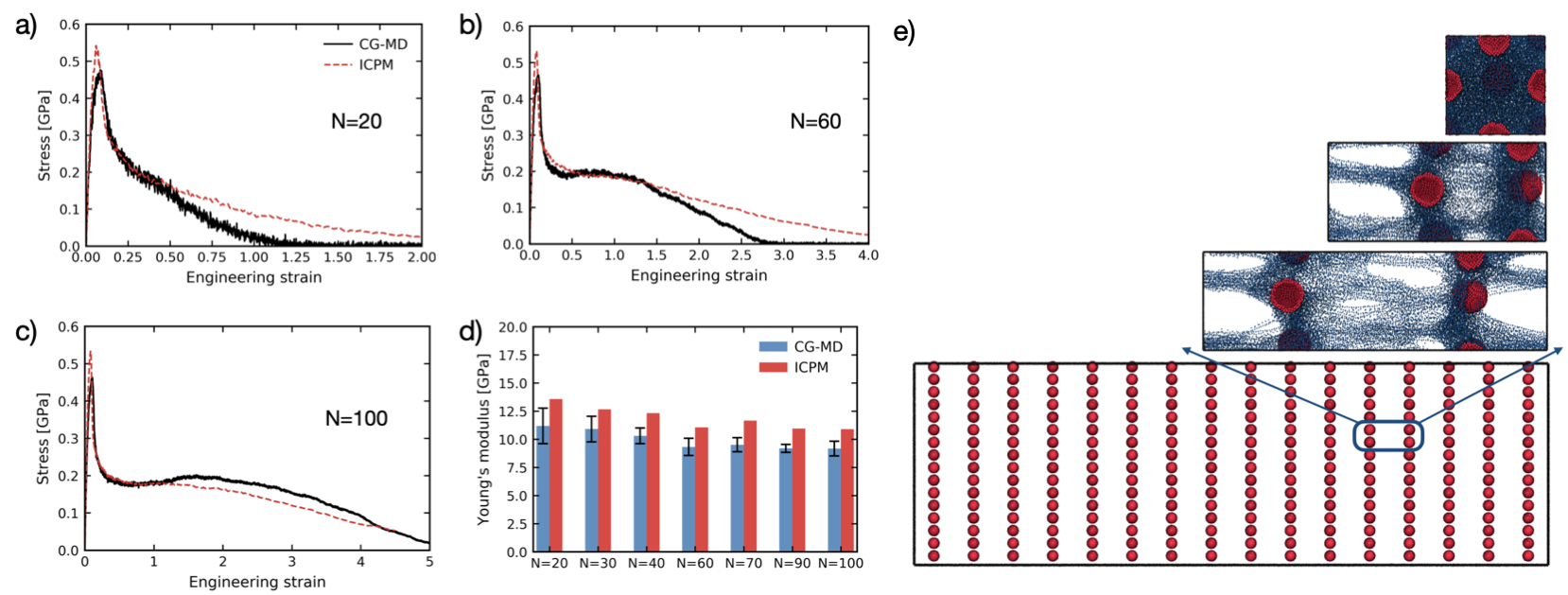}
    \caption{Stress-strain curves from uniaxial tensile deformations of CG-MD (black dashed lines) and ICPM models (red solid lines) of FCC structures of PGNs with PMMA grafts a) $N=20$ monomers,  b) $N=60$ monomers and  c) $N=100$ monomers; d) Young's modulus of PGNs with various PMMA graft lengths from CG-MD (blue) and ICPM models (red); e) Schematic for the uniaxial tensile deformation of PMMA-PGN with graft chain length $N=100$ monomers in CG-MD and ICPM representations.}
    \label{fig:uniaxial_deform_stress_modulus}
\end{figure}

The PMF used in the ICPM model is derived from dilation and compression, while its transferability to other mechanical loadings requires further  examination. Here, we test the ICPM model for uniaxial deformation of matrix-free PGNs with varying graft chain lengths. Snapshots of the uniaxial deformation of matrix-free PGNs in CG-MD and ICPM representations can be found in Figure \ref{fig:uniaxial_deform_stress_modulus} (e). Figure \ref{fig:uniaxial_deform_stress_modulus} (a)-(c) present the stress-strain curves for matrix-free PGNs with graft length $N=20$, $N=60$, and $N=100$ monomers from both ICPM and CG-MD simulations. Compared to the corresponding CG-MD simulations, the ICPM model generally captures the stress-strain behavior during the uniaxial tensile deformation, but has a slight overestimation of the maximum stress for all graft chain lengths. Furthermore, we calculate Young's modulus for all of these systems, as shown in Figure \ref{fig:uniaxial_deform_stress_modulus} (d). Similarly to the bulk modulus in PGNs, Young's modulus decreases with increasing graft chain length, consistent with measurements in a recent experimental work\cite{jhalaria_unusual_2022}. Specifically, increasing the volume fraction of the polymer, which is the softer phase, results in a steady decrease in elastic constants with increasing $N$. Although the modulus evaluated from the ICPM models is consistently $\sim 15\%$ higher than the corresponding CG-MD results at each graft chain length, the general trends for the N-dependence of the modulus are adequately captured, implying that the transferability of the PMF obtained from dilation/compression loadings to uniaxial tensile deformation is satisfactory. Future work may further examine whether PMFs in the ICPM model can be further optimized for various mechanical loading scenarios. This will inevitably complicate the force field development effort and may be explored in future studies. We also note that the uniaxial tensile deformation is carried out with a fixed cross-section area to mimic the local craze zones in polymers undergoing failure as is done in previous studies\cite{rottler_growth_2003}, to capture the triaxiality of the materials' stress-state near the crack tip. As such, the material constants such as Young's modulus may differ slightly from macroscale uniaxial tension experiments. 

\begin{figure}
    \centering
    \includegraphics[width=1.0\textwidth]{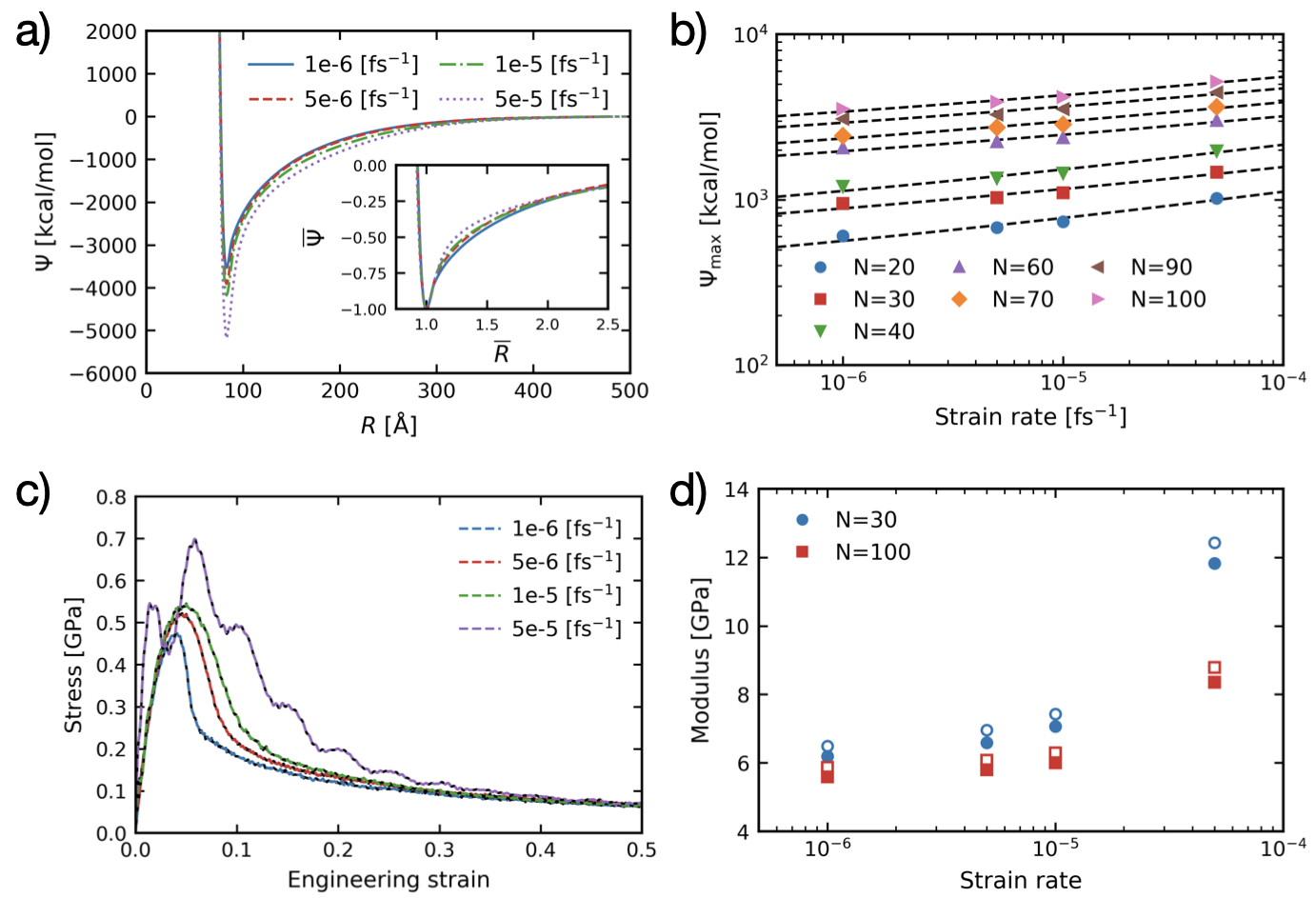}
    \caption{(a) PMF derived from triaxial tensile deformation of CG-MD models under strain rates $\dot{\gamma}=1e-6$ $\mathrm{fs^{-1}}$ (blue), $\dot{\gamma}=5e-6$ $\mathrm{fs^{-1}}$ (red), $\dot{\gamma}=1e-5$ $\mathrm{fs^{-1}}$ (green), and $\dot{\gamma}=5e-5$ $\mathrm{fs^{-1}}$ (purple); (b) Well depth of the PMF as a function of strain rates for PMMA-PGN with various graft chain length; (c) Stress-strain curves from triaxial tensile deformation of CG-MD (solid lines) and ICPM (dashed lines) model under various strain rates; (d) Bulk modulus estimated from CG-MD (filled symbols) and ICPM (hollow symbols) as a function of strain rates.}
    \label{fig:strain_rate}
\end{figure}

\subsection{Strain-Rate Dependence}
The simulations presented so far are performed at a constant strain rate $\dot{\epsilon}=1 \times 10^{-6}$ $\mathrm{fs^{-1}}$. However, the strain rate does affect the mechanical properties of polymeric materials due to their viscoelastic nature. It is practically meaningful for the proposed method to capture this mechanism. We first examine the PMF $\Psi(r,\dot{\epsilon})$ derived from the CG-MD models at different strain rates from $\dot{\epsilon}=1\times 10^{-6}$ to $\dot{\epsilon}=5\times 10^{-5}$ $\mathrm{fs^{-1}}$ for PMMA-PGNs with grafts $N=100$ monomers, as shown in Figure \ref{fig:strain_rate} (a). Clearly, PMFs $\Psi(r,\dot{\epsilon})$ vary with strain rates $\dot{\epsilon}$. This is expected given that the work function extracted from bulk dilation /compression has entropic contributions and includes dissipative processes at non-equilibrium rates. For example, the apparent depth of the potential well increases as the strain rate increases $\dot{\epsilon}$. It indicates that the PMF derived from the single state cannot simply be transferred between different mechanical loading rates. Nevertheless, the inset in Figure \ref{fig:strain_rate} (a) shows that the PMFs at different strain rates collapse when normalized by the equilibrium distance between the particles and the apparent depth of the potential well $\Psi_{\mathrm{max}}$. Therefore, it is possible to capture the rate dependence of PMFs by changing the apparent depth of the potential well $\Psi_{\mathrm{max}}$. 

In Figure \ref{fig:strain_rate} b), the apparent depth of the potential well $\Psi_{\mathrm{max}}$ is plotted as a function of the strain rates for various PMMA graft lengths. We find that $\Psi_{\mathrm{max}}(N)$ increases as the strain rate $\dot{\epsilon}$ increases for all lengths of the graft chain examined in this work. The Cowper-Symonds (C-S) model, which is widely used to describe strain-rate dependent material properties, is applied to fit the apparent depth dependent on the strain rate of the potential well $\Psi_{\mathrm{max}}(N)$. The formula for the C-S model is given below:
\begin{equation} 
   \Psi_{\mathrm{max}}(\dot{\epsilon})=\Psi_{0}(1+(\frac{\dot{\epsilon}}{\dot{\epsilon}_0})^n)
\end{equation}
where $\Psi_{0}$ denotes the zero rate limit of the apparent depth of the potential well; $\dot{\epsilon}_0$ represents a critical strain rate, below which (e.g., $\dot{\epsilon}/\dot{\epsilon}_0\ll 1$) the strain rate $\dot{\epsilon}$ has a negligible effect. The power-law exponent $n$ describes the degree of non-Newtonian behavior in the high strain rate regime, where $\dot{\epsilon}/\dot{\epsilon}_0\gg 1$. It is noted that the critical strain rate $\dot{\epsilon}_0$ and the power-law exponent $n$ are typically intrinsic constants dependent on the material. We find that $\dot{\epsilon}_0$ is not quite sensitive to the graft chain length, and it is fixed at $\dot{\epsilon}_0=1\times 10^{-7}$ $\mathrm{fs^{-1}}$ during all fittings. This critical value that we obtained is in line with the time scale of segmental relaxation of PMMA at $T=300$ K, similar to our previous finding\cite{hansoge_effect_2019}. The power-law exponent is observed to be slightly dependent on graft chain length, i.e., from $\sim 0.2$ for long to $\sim 0.4$ for short grafts, the range of which is indeed within the reported values of simulations and experiments\cite{meng_crazing_2016,hansoge_universal_2021}. The stress-strain curves of different strain rates from ICPM simulations of PGNs with PMMA grafts $N=100$ monomers are shown in Figure \ref{fig:strain_rate} c). With PMFs derived and optimized at respective strain rates, the ICPM model can successfully capture stress-strain behaviors at all strain rates tested. Furthermore, Figure \ref{fig:strain_rate} d) shows that the bulk modulus estimated from the corresponding stress-strain curves increases with increasing strain rate in ICPM simulations of two example ICPM PGN systems with graft chain length $N=30$ and $N=100$, which are in good agreement with the reference CG-MD.

\subsection{Density and Structure of Matrix-Free PGNs}

\begin{figure}
    \centering
    \includegraphics[width=1.0\textwidth]{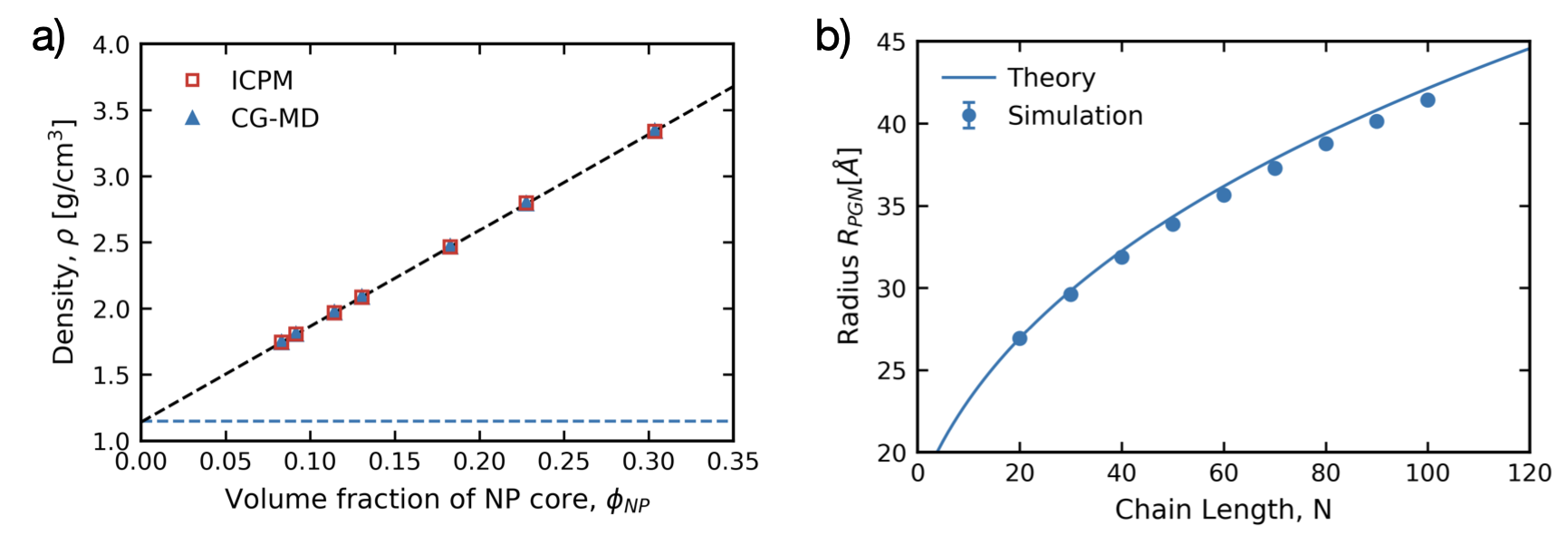}
    \caption{(a) Comparison of the density $\rho$ as a function of volume fraction of nanoparticle core $\phi_{NP}$ for the CG-MD (blue triangles) and ICPM (red squares) models of PMMA-PGN melts. The dashed blue line is the reference density from the CG-MD model of PMMA melts at $T=300$ K. The dashed black line is the linear fit of the ICPM values. (b) Radius of a single PGN as a function of graft chain length computed from our simulation (symbols) and the theoretical model (solid line) from Midya \textit{et al.}\cite{midya_structure_2020}}
    \label{fig:Density}
\end{figure}

Density is one of the most important thermodynamical properties for soft materials. Here, we examine the performance of the ICPM model of matrix-free PGNs in terms of the reproduction of the density of the reference CG-MD models. Figure \ref{fig:Density} (a) illustrates the densities of matrix-free PGNs with various graft chain lengths as a function of the volume fraction of the NP core from ICPM and CG-MD simulations. It is observed that the ICPM model can quantitatively reproduce the density of the CG-MD models in the respective volume fractions of the NP cores $\phi_{NP}$. In addition, we find that a linear function can be successfully applied to fit these results, and the extrapolation to the zero volume fraction limit of the NP cores is in good agreement with the reference density of the pure PMMA melt at $T=300$ K\cite{hsu_systematic_2014}. Moreover, we compare the radius of the PGN calculated from our simulations with a recent theoretical model, namely, two-layer model, developed by Midya \textit{et al.}\cite{midya_structure_2020} to further validate our simulation models. The radius of the PGN particle is defined as: $R_{PGN}=R_{NP}+h$, where $R_{NP}$ is the radius of the NP core and $h$ is the thickness of the polymer layer. In our simulations, the radius of the PGN is calculated from the position of the first peak in the radial distribution functions of the PGNs. This value essentially corresponds to the nearest-neighbor distance between the PGNs. The radius of a PGN predicted by the two-layer model is calculated following Equation $S16$ from Midya et al. as \cite{midya_structure_2020}:
\begin{equation}
    R_{PGN}^{theo}=\big(\frac{\pi}{\sqrt{18}}\big)^{1/3}(R_{NP}^3+\frac{3Z N}{4\pi \rho_{poly}})
\end{equation}
where $Z=4\pi R_{NP}^2 \sigma$, and the polymer density $\rho_{poly}$ is obtained from the corresponding CG-MD simulations, considering the fact that the shape of the PGN in the dense state follows the packing of Voronoi cells. Figure \ref{fig:Density} (b) shows that the simulation results of the radius of the PGN are in good agreement with the theoretical predictions.

%\begin{figure}
%    \centering
%        \includegraphics[width=1.0\textwidth]{figures/theo_NP_space.png}
%    \caption{(a) Schematic of the two-layer model for PGNs proposed by Midya \textit{et al.} \cite{midya_structure_2020} (b) Comparison of the PGN radius from ICPM model and two-layer models as a function of graft langth N: blue symbols are the radius of the PGN computed from the ICPM with the dashed line being an exponential fit; solid and dotted lines represent total radius $R_{tot}$ and the radius of core size and dry layer $R_{core}+h_{hry}$ of a PGN in the two-layer model, respectively; The grey solid lines are linear lines to show the distint regimes, and the vertical dashed line is the critical length $N_{cr}$ of the transition from a concentrated brush regime to a semidilute brush regime\cite{hansoge_effect_2019}.  }
    \label{fig:theo_NP_space}

\subsection{Computational Efficiency}
\begin{figure}
    \centering
    \includegraphics{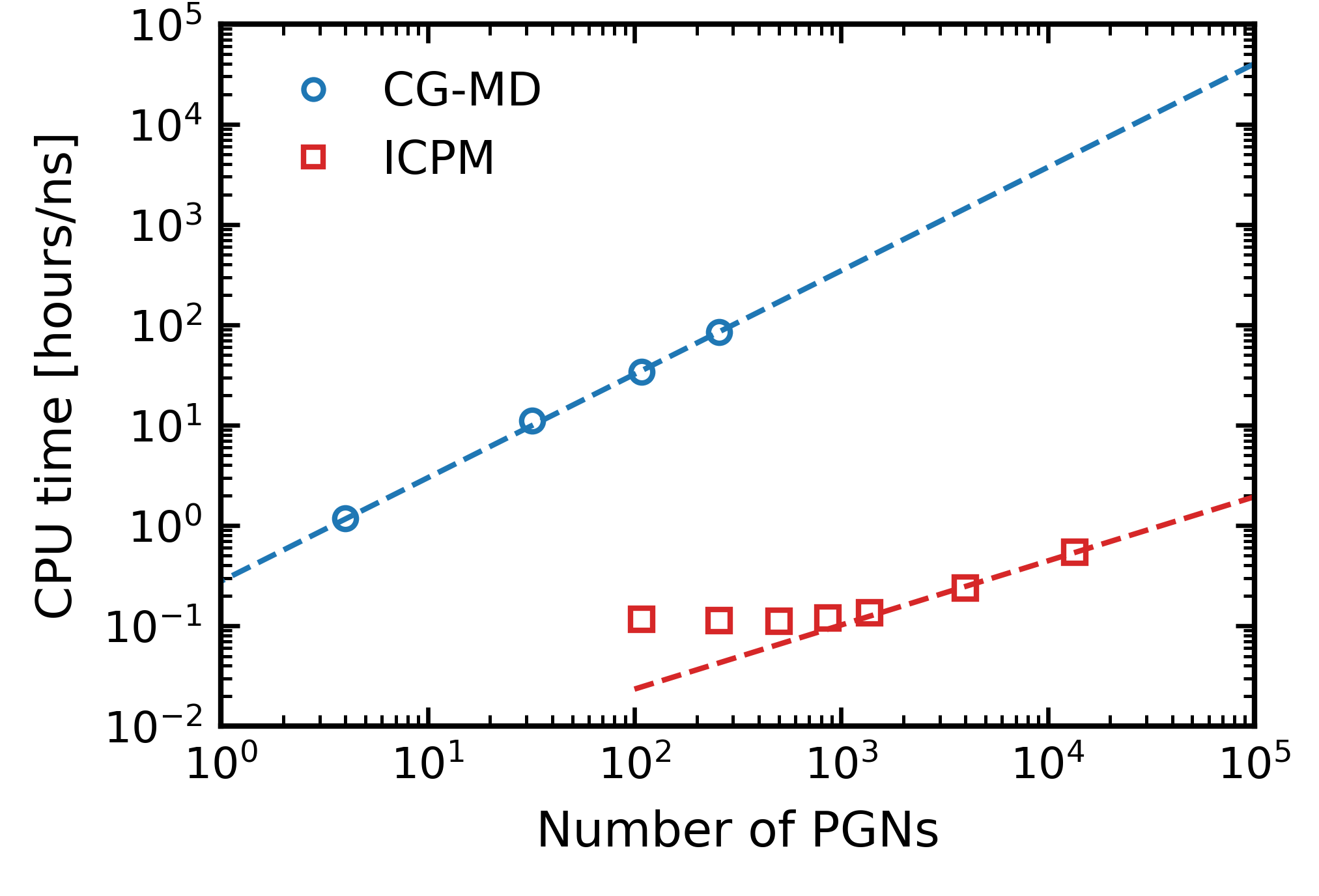}
    \caption{Comparison of the CG-MD (blue) and ICPM (red) PMMA-PGN model with 10 ps simulation time utilizing 24 CPU cores. The dashed lines represent the fitting lines to the data via the equations $y=2.8\times10^{-1}\times x^{1.03}$ and $y=1.2\times10^{-3}\times x^{0.64}$ for CG-MD and ICPM systems, respectively.}
    \label{fig:speed_particles_CPU}
\end{figure}

One of the most important advantages of coarse graining is the speed-up in computation. To demonstrate the computational acceleration of the proposed ICPM model, we choose benchmark systems composed of various numbers of PGNs in both the CG-MD and ICPM representations. These benchmarked PMMA-PGN systems have fixed graft length $N=100$ monomers and density $\sigma=0.5$ $\mathrm{chains/nm^2}$, which have a total of 6059 and 1 simulated beads per PGN particle in the CG-MD and ICPM representations, respectively. Reducing the degree of freedom can significantly accelerate ICPM simulations. We estimate the CPU time required to run 1 ns simulations from $10$ picoseconds of both ICPM and CG-MD simulations with the same time unit $\delta t=0.001$ femtosecond using 24 cores of one or two of these processors that have similar performance connected with Infiniband (Intel Xeon E5-2680, v4 or Intel Xeon Gold 6132 or Intel(R) Xeon(R) Gold 6230). As shown in Figure \ref{fig:speed_particles_CPU}, the CPU time of the CG-MD model follows a linear trend as the number of PGN increases on double logarithmic scales. In ICPM simulations, there is a plateau in the regime where the number of PGNs $N_{PGN}<1000$ due to the inefficiency of parallelization for a few PGNs. It increases linearly with the number of PGNs $N_{PGN}>1000$. The computational speed-up of the ICPM model compared to CG-MD is more than the multiplication of a constant. The power-law exponent of the required CPU time as a function of the number of PGNs even decreases in the ICPM model. As a consequence, the ICPM model can gain enormous acceleration for systems with a large number of PGNs. For example, it is approximately 5 orders of magnitude faster for the ICPM model than for the CG-MD model when simulating $\sim 1\times 10^5$ PGN particles in a cube with edge length $\sim 0.5$ $\mathrm{\mu m}$.

\section{Discussion and Conclusion}

In this contribution, we propose a hierarchical coarse-graining framework to construct an efficient particle-based model for simulations of matrix-free PGNs. In this highly coarse-grained model, i.e., ICPM, the polymer grafts are implicitly modeled by effective interactions derived using the strain-energy mapping approach, leading to a single-particle representation of an entire PGN. As a first attempt towards such an aggressive coarse-grained model, we employ a simple pairwise form of effective interactions to represent the potential energy surface. Large-scale ICPM molecular dynamics simulations of matrix-free PGNs with these effective pairwise potentials are performed. We demonstrate the effectiveness of the ICPM method by performing various mechanical loadings, e.g., bulk dilation, compression, and uniaxial tensile tests, and comparing the predicted material properties with the finer-grained CG-MD simulations. In general, the ICPM model can accurately predict multiple thermodynamic and mechanical properties, including density, stress-strain curves, Young's modulus, bulk modulus, and toughness. In the ICPM simulations, we observe that as the graft chain length increases, the bulk and Young's modulus of matrix-free PGNs decrease. This happens because the volume fraction of the NP core decreases as the graft length increases, and the contribution of the NP core that has a higher modulus decreases. This observation is indeed consistent with recent experimental measurements of the bulk modulus as a function of the volume fraction of the NP cores\cite{jhalaria_unusual_2022}, as seen in Figure \ref{fig:bulk_modulus_toughness} (c). Additionally, the bulk toughness of matrix-free PGNs increases with increasing graft length, which is consistent with our previous work on polymer grafted slabs\cite{hansoge_universal_2021}. Recent studies suggested that chain entanglements, particularly interparticle entanglements, play an important role in the fracture and toughness of matrix-free PGNs\cite{ethier_structure_2018,ethier_uniaxial_2019}. Since the longest chain length tested in the current work is 100 monomers, approaching the critical entanglement length of PMMA ($\sim 140$ monomers\cite{graessley_entanglement_1974}), our work applies to unentangled systems and leaves understanding of the mechanisms of topological constraints in the mechanical properties of matrix-free PGNs to future studies. Moreover, the extraction of the ICPM model from a fine-grained model relies on a unit cell based on a closed-pack configuration assumption. We revisited this assumption and examined how the properties would change in the case of amorphous structures. Recent experimental studies found that distinct PGN structures can be obtained by varying processing conditions\cite{heo_thermally_2013,liu_microscopic_2019}, increasing the complexity for systematic investigations. For benchmarking purposes, we employ a simple annealing protocol and test the performance of ICPM on simulating amorphous matrix-free PGNs. In brief, an amorphous state of PGNs can be obtained by sufficient annealing from a low-density PGN system in ICPM simulations. Details of the simulation and mechanical tests on matrix-free PGNs at amorphous state are given \textbf{Supporting Information}.

The ICPM model is capable of capturing a wide range of responses of the strain rate on the mechanical properties of matrix-free PGNs. Our results show that the PMF deepens with increasing strain rates. This rate-dependent PMF can be estimated without conducting fine-grained simulations using PMF at a given strain rate using the universality of the rate-dependent PMFs that overlaps when normalized by the equilibrium distance between particles and the well depth of the PMF. This will contribute to a better understanding of the effects of strain rate on the mechanical properties of polymer composites\cite{jacob_strain_2004}. Notably, this capability makes our ICPM model differ from most other potential-of-mean-force based models, e.g., where free energy perturbation\cite{masoumi_nanolayered_2020} or Boltzmann inversion\cite{pan_interaction_2022} is used to estimate the PMF. Additionally, the ICPM model accurately captures thermodynamic properties such as the density of matrix-free PGNs with various graft chain lengths. The interparticle distance and PGN radius can also be quantitatively evaluated from the ICPM simulations. The predicted quantities agree well with a recent theoretical model\cite{midya_structure_2020}.

The computational efficiency of the ICPM model is demonstrated by a speed-up of $\sim 5$ orders of magnitude compared to the reference CG-MD model. The single-particle representation of the PGN particle in the ICPM model significantly reduces the degree of freedom, allowing direct simulations of the micronscale system ($\sim 1\times 10^5$ particles). This capability will enable the study of, e.g., self-assembly, fracture behavior, and microballistic impact tests. These properties are practically inaccessible to CG-MD models of matrix-free PGNs\cite{hansoge_materials_2018,midya_structure_2020}. Furthermore, the idea of ICPM that implicitly models grafts with effective interactions is, in principle, applicable not only to PGNs but also to colloidal nanocrystals with arbitrary geometry and biomolecular grafts such as DNA strands\cite{wang_emergence_2022}. Recently, an experimental calibration of effective interactions between DNA-coated colloids was developed to understand their melting and crystallization at the microscopic level. Combining these theoretical models with our approach for bottom-up derivation of PMFs is encouraging for more efficient and accurate multiscale modeling of superlattices of ligand-coated colloids\cite{boles_self-assembly_2016} and the matrix-free PGNs. 

The rate dependence captured by the ICPM model paves the way for high-strain rate fracture behavior and impact resistance of matrix-free PGNs. This makes it necessary to investigate the predictive power of the ICPM model for other types of dynamic loading conditions, such as shear deformation. In fact, because of the assumption of short-range NN pair-wise interactions, large-scale flows of matrix-free PGNs are not anticipated to be accurately modeled in the current ICPM approach. Future iterations of the ICPM upscaling approach may benefit from fitting to a broad range of mechanical tests to calibrate potentials. The model transferability may require better representations of the potential energy surface to account for many-body effects using, e.g., machine-learning force-fields that are able to incorporate local environments such as radial and angular interactions\cite{unke_machine_2021}. In addition, recent advances in studying the self-assembly of multicomponent superlattices suggest that the introduction of dispersity can provide enormous possibilities for exploring the phase diagrams of ligand (e.g., polymer) coated nanoparticles\cite{lacour_tuning_2022}. Understanding and validating how dispersity can be utilized to reach uncharted regions of the material design space for improved mechanical properties will be important and meaningful for the rational design of matrix-free PGNs with target performance in the future. 

\begin{acknowledgement}

%Please use ``The authors thank \ldots'' rather than ``The
%authors would like to thank \ldots''.

The authors acknowledge funding from the Center for Hierarchical Materials Design at Northwestern University and Army Research Office. A supercomputing grant from Quest High-Performance-Computing System at Northwestern University is also acknowledged.
\end{acknowledgement}

\begin{suppinfo}
Force-fields of Poly(methyl methacrylate) and nanoparticles; Simulation details of ICPM model of matrix-free PGNs; Simulation details and equilibration protocol of CG-MD models; Polymer conformations in CG-MD simulations; Discussions on ICPM simulations of amorphous PMMA-grafted nanoparticles including the annealing protocol and structure and mechanical properties of the amorphous matrix-free PGNs. 

%Size effect at the ICPM level

\end{suppinfo}

\bibliography{manuscript}

\end{document}